\documentclass[english,tightenlines,eqsecnum,amsmath,amssymb,aps,prd,superscriptaddress,nofootinbib,showpacs,floats]{article}
\usepackage{amsmath,amsfonts,amssymb,multicol}
\DeclareMathOperator{\arcsinh}{arcsinh}
\usepackage{graphicx,caption,subcaption}
\usepackage{epstopdf}
\usepackage{enumitem} 
\usepackage[usenames]{xcolor}
\usepackage{epstopdf}
\definecolor{AHZ}{rgb}{0.0,1,0.0}
\usepackage[colorlinks,linkcolor=AHZ,citecolor=red,bookmarks]{hyperref}
\usepackage{rotating}
\usepackage[mathscr]{eucal}
\usepackage{graphicx}
\usepackage[T1]{fontenc}
\usepackage{babel}
\usepackage{authblk}
\usepackage{epsfig}
\usepackage{color}
\usepackage{amssymb}
\usepackage{fancyhdr}

\def\nn{\nonumber\\}
\newcommand{\f}[2]{\frac{#1}{#2}}
\def\be{\begin{equation}}
\def\ee{\end{equation}}
\def\bea{\begin{eqnarray}}
\def\eea{\end{eqnarray}}

\def\bwt{\begin{widetext}}
	\def\ewt{\end{widetext}}

\usepackage{amsmath}

\usepackage{amssymb}

\begin{document}
	
	\title{Wormhole solutions in generalized Rastall gravity}
	\author{Naser Sadeghnezhad\thanks{nsadegh@maragheh.ac.ir}}
	\affil{{\rm Research~Institute~for~Astronomy~and~Astrophysics~of~ Maragha~(RIAAM), University of Maragheh,  P.~O.~Box~55136-553,~Maragheh, Iran}}
	\renewcommand\Authands{ and }
	\maketitle
\begin{abstract}
Our aim in the present study is to search for static spherically symmetric wormhole solutions in generalized Rastall gravity (GRG). In this theory, a varying coupling parameter can be regarded as the source of dark energy (DE) and hence, responsible for the current accelerated expansion of the Universe. We assume an anisotropic energy momentum tensor (EMT) with a linear relation between energy density and pressures profiles, i.e., $p_r(r)=w_1\rho(r)$ and $p_t(r)=w_2\rho(r)$, and obtain two classes of solutions to the field equations of GRG, including the solutions with zero and nonzero redshift functions. For these solutions we find that the matter distribution obeys the physical reasonability conditions, i.e., the flare-out, the weak (WEC) and null (NEC) energy conditions either at the throat and beyond it. These conditions put restrictions on model parameters, e.g. the equation of state (EoS) parameters and the value of energy density at the throat. Hence, we find that in the framework of GRG, asymptotically flat wormhole configurations can be built without resorting to exotic matter. We further study properties of timelike as well as null geodesics for the obtained wormhole solutions. In the case of null geodesics, gravitational lensing effects are discussed and it is shown that the throat of wormhole can effectively act as a photon sphere near which the light deflection angle takes arbitrarily large values.
\end{abstract}
\maketitle
\section{Introduction}
Conceptually, wormholes can be imagined as tube-like geometric structures connecting two separate regions within the same Universe or even different Universes altogether. The idea of such interesting geometries can originally be traced back to Ludwig Flamm~\cite{Flamm} in 1916. He found that Schwarzschild solution can represent what we now know as a wormhole and in the 1920s, Herman Weyl speculated possible roles of Schwarzschild solution and other wormholes in physics~\cite{Weyl1920}. In 1935, Einstein and Rosen tried to obtain models for elementary particles which represent a bridge that provides a link between two identical sheets. The result of their work is what is now known as the Einstein-Rosen bridge~\cite{ERBR}. Historically, the term \lq{}\lq{}wormhole\rq{}\rq{} was first introduced by Wheeler in 1957~\cite{Wheel1972} and further developed by Misner and Wheeler~\cite{MisWheel,Wheelerworm}. However, it was later realized that Wheeler wormholes are not traversable since they do not allow a two way communication between two regions of the spacetime through a minimal surface called the wormhole throat~\cite{FulWheel}. 
\par
Over the past decades, there has been a huge amount of interest in probing the physical properties of wormhole structures. The main motivation for such interest comes from the seminal papers by Morris, Thorne and Yurtsever~\cite{mt,mt1}. Having introduced a static spherically symmetric metric, the authors of thees articles investigated Lorentzian traversable wormholes and developed physical circumstances and key properties for such objects that allow a traveler to experience crossing through the throat of wormhole back and forth in both Universes. In the framework of classical general relativity (GR) matter fields threading wormhole structures play a significant role. But, it was found that traversable wormholes possess an EMT that violates the null energy condition (NEC)~\cite{mt},\cite{khu} which is a fundamental condition in GR. This condition states that for any null path, the energy density of the matter and fields must be non-negative. As the NEC is the weakest among all the classical energy conditions, its violation implies that the other energy conditions such as WEC are also violated. In the case of traversable wormholes, the type of matter that would allow for the existence of such structures is referred to as exotic matter. Such type of matter is characterized by a negative energy density e.g., ghost scalar fields or phantom energy~\cite{ghostsfph}, and consequently involves an EMT that violates the NEC. In this sense, while the idea of traversable wormholes opens up fascinating possibilities for interstellar travel via connecting distant regions of spacetime~\cite{mt,mt1} and time machines~\cite{timemach}, it also challenges our understanding of the fundamental laws of physics and the nature of matter throughout the Universe. Generally, it is believed that
classical matter fields respect energy conditions~\cite{HawCMF}, but in fact, it has been shown that they can be violated by some quantum fields e.g., Casimir effect and Hawking evaporation~\cite{Caseffect}. Moreover, it has been shown that wormhole geometry with negative energy density may be created by gravitational squeezing of the vacuum~\cite{negendensqueez}, see also~\cite{khu,Klinkhammer1991} for more details.
\par
Research in wormhole structures became active in the last decades since the work of Morris, Thorne, and Yurtsever~\cite{mt,mt1}. Numerous and extensive works along this line have provided a new field of study in theoretical physics and several publications have been reported in the literature in recent years, for example, static spherically symmetric wormholes supported by phantom energy or ghost scalar fields~\cite{ghostsfph},\cite{phantworm}, wormhole solutions with Casimir energy as the source~\cite{mt1,csworm} and interacting DE and dark matter~\cite{intdarksec}, see~\cite{lobocgqreview} for recent reviews. However, since exotic matter sources violate energy conditions, many attempts have been made so far toward avoiding or at least minimizing the usage of such type of matter fields and instead, modifying GR with the purpose of overcoming the issue of energy conditions within wormhole structures. In this regard, a great deal of time and effort has gone into investigating wormhole solutions in different modified gravity theories such as, scalar-tensor theories~\cite{bd}, modified gravities with curvature-matter coupling~\cite{Garcia-Lobo}, Lovelock theories of gravity~\cite{LOVEWORM}, $f(R)$ gravity~\cite{fr}, Einstein-Cartan theory~\cite{ectwormhole1,ectwormhole}, Einstein-Gauss-Bonnet~\cite{gmfl}, brane-world scenarios~\cite{branww} and other theories~\cite{otherworms}. The study of wormhole structures has been also done in Rastall gravity~\cite{wormrastrace,morad1,morad2}. This theory was introduced as a modification to GR by coupling the matter field with spacetime geometry in a non-minimal way~\cite{ras}. In this setup, Rastall argued that the ordinary conservation law of EMT, i.e., $\nabla_{\mu}T^{\mu\nu}=0$ may no longer hold in curved spacetime. The basic motivation for such argument is that the conservation laws have
been only tested on the Minkowski spacetime or quasistatic gravitational fields~\cite{morad3}. Another justification for considering modified theories which violate the EMT conservation is to provide a setting to discuss the process of particle production~\cite{ppp}. However, it is known that the conservation of EMT is not obeyed by the particle production process~\cite{ppp1}. Thus, Rastall modified this usual conservation law as, $\nabla_{\mu}T^{\mu\nu}=\lambda\nabla^\nu R$, i.e., the conservation law is proportional to the gradient
of the Ricci scalar~\cite{ras}. Since its advent, Rastall gravitational theory has attracted the attention of many researchers due to its potential applications in cosmological
and astrophysical scenarios~\cite{rascosastro}. This theory is also in good agreement with various observational data and theoretical expectations~\cite{morad4}. 
\par
Recently, a generalization of Rastall gravity has been introduced in which the Rastall parameter, $\lambda$, which is a measure of mutual interaction between matter and geometry, is a varying parameter that depends on spacetime coordinates~\cite{morad3}. It has been argued that in this generalized Rastall gravity (GRG) a running coupling between matter and geometry can play the role of dark energy and hence the current accelerated expansion phase of the
Universe phase may have been originated from such a dynamic matter-geometry interaction. Moreover, the authors of~\cite{morad3} have shown that in a flat FLRW background, GRG is
capable of explaining an inflationary phase even without a matter component. Also, the existence of pre- and post-inflationary solutions in GRG has been reported in~\cite{GRG22} and in~\cite{ZiaRas}, it has been shown that a varying Rastall parameter could prevent singularity formation at the end of a collapse scenario. Motivated by the above considerations, our aim in the present work is to study wormholes solutions in the framework of GRG. The paper is then organized as follows: In Sec.~(\ref{FES}) we briefly review the field equations of GRG. In Subsecs.~(\ref{phizero}) and (\ref{phinzero}) we try to find exact solutions representing zero as well as non-zero tidal force wormhole structures. Sec.~(\ref{glens}) is dedicated to studying geodesic equations for timelike particles and null rays, along with gravitational lensing features of the obtained wormhole solutions. Our conclusions are drawn in Sec.~(\ref{concluding}).
\par
\section{Field equations of GRG and Wormhole Solutions}\label{FES}
In the framework of original Rastall theory, the Rastall parameter which is a measure of coupling between matter and curvature is a constant~\cite{ras}. However, this theory cannot remove the need to a mysterious dark energy source in order to describe the accelerated expansion of the Universe~\cite{rascosmiceras1}. In this sense, a simple generalization of Rastall theory has been proposed as
\be\label{CoVDiv}
\nabla_{\mu}T^{\mu}_{\,\,\,\nu}=\nabla_{\nu}(\lambda R),
\ee
in which the Rastall parameter is no longer a constant and can describe the accelerated Universe without invoking to a dark energy source~\cite{morad3}. The field equations are as follows
\be\label{genrasfieldeqs} 
G_{\mu\nu}+\kappa\lambda g_{\mu\nu}R=\kappa T_{\mu\nu},
\ee 
where $\kappa$ is Rastall gravitational constant \cite{morad3}.
\par
Let us now consider the general static and spherically symmetric line element representing a wormhole spacetime given by
\begin{eqnarray}\label{evw}
ds^2=-{\rm e}^{2\Phi(r)}dt^2+\left(1-\f{b(r)}{r}\right)^{-1}dr^2+r^2d\Omega^2,
\end{eqnarray}
where $d\Omega^2=d\theta^2+\sin^2\theta d\phi^2$ is the standard line element on a unit two-sphere, $\Phi(r)$ is the redshift function and $b(r)$ is the wormhole shape function. The radial coordinate ranges from $r_0$ (wormhole\rq{}s throat) to spatial infinity. At the throat, defined by the condition $r_0=b(r_0)$, there is a coordinate singularity where the radial metric component ${\sf g}_{rr}$ diverges, however, the radial proper distance
\be\label{radpropdis}
\ell(r)=\pm\int_{r_0}^{r}\f{dx}{\left(1-b(x)/x\right)^{1/2}},
\ee
is required to be finite. Indeed, at the throat we have $\ell(r_0)=0$ while, $\ell<0 (>0)$ on the left (right) side of the throat. Conditions on redshift and shape functions under which, wormholes are traversable have been discussed completely in~\cite{mt}. Traversability of the wormhole requires that the spacetime be free of horizons which are defined as the surfaces with ${\rm e }^{2\Phi(r)}\rightarrow0$; therefore the redshift function must be finite everywhere. The shape function should satisfy the following conditions~~\cite{mt,mt1}:
\begin{enumerate}[label=(\roman*)]
\item $\left(1-\f{b(r)}{r}\right)\Big|_{r=r_0}=0$: This gives the radius of the throat at $r=r_0$.\label{c1}
\item $rb^\prime(r)-b(r)<0$ for $r>r_0$: This is the fundamental flare-out condition which at the throat reads, $b^\prime(r_0)<1$.\label{c2}
\item $\f{b(r)}{r}<1$ for $r>r_0$: This condition guarantees that the (Lorentzian) metric signature is preserved.\label{c3}
\item $\lim_{r\rightarrow\infty}\f{b(r)}{r}=0$: This condition gives the asymptotic flatness of the metric.\label{c4}
\end{enumerate}	
 Let us now define the time-like and space-like vector fields, respectively as $u^i=[1,0,0,0]$ and $v^i=\left[0,\sqrt{1-b(r)/r},0,0\right]$, so that $u^iu_i=-1$ and $v^jv_j=1$. The anisotropic EMT of matter source then takes the form
\be\label{emtaniso}
T_{ij}=[\rho(r)+p_t(r)]u_iu_j+p_t(r)g_{ij}+[p_r(r)-p_t(r)]v_iv_j,
\ee
with $\rho(r)$, $p_r(r)$, and $p_t(r)$ being the energy density, radial and tangential pressures, respectively. We therefore obtain the components of field equations (\ref{CoVDiv}) and (\ref{genrasfieldeqs}) as 
\bea
&&\!\!\!\!\!\!\!\!\!\!\!\!\!\!\!\!\!\!\!\!2r^2\lambda(r)(r-b(r))\Phi^{\prime\prime\prime}(r)+4r\left(r\lambda(r)(r-b(r))\Phi^\prime(r)+\f{r}{2}\lambda^\prime(r)(r-b(r))-\f{3\lambda(r)}{4}\left[rb^\prime(r)-\f{4}{3}r+\f{b(r)}{3}\right]\right)\Phi^{\prime\prime}(r)\nn&-&r\lambda(r)\left(r\Phi^{\prime}(r)+2\right)b^{\prime\prime}(r)+2r\left[r(r-b(r))\lambda^\prime(r)-\lambda(r)(rb^\prime(r)-b(r))\right]\Phi^{\prime2}(r)\nn&+&\left[-r\lambda^\prime(r)(3b(r)-4r+rb^\prime(r))-2r\lambda(r)b^\prime(r)+\lambda(6b(r)-4r)+r^3(p_r(r)+\rho(r))\right]\Phi^\prime(r)\nn&-&2rb^\prime(r)\lambda^\prime(r)+4\lambda(r)b^\prime(r)+r^2\left[rp_r^\prime(r)+2p_r(r)-2p_t(r)\right]=0,\label{consexp}\\
\rho(r)&=&2\lambda(r)\left(1-\f{b(r)}{r}\right)\left[\Phi^{\prime\prime}(r)+\Phi^{\prime2}(r)\right]-\lambda(r)\left[\f{b^\prime(r)}{r}-\f{4}{r}+\f{3b(r)}{r}\right]\Phi^{\prime}(r)+(1-2\kappa\lambda(r))\f{b^\prime(r)}{\kappa r^2},\label{rhoex}\\
p_r(r)&=&-2\lambda(r)\left(1-\f{b(r)}{r}\right)\left[\Phi^{\prime\prime}(r)+\Phi^{\prime2}(r)\right]+\left[\f{\lambda(r)b^\prime(r)}{r}-\f{4\lambda(r)}{r}\left(1-\f{3b(r)}{4r}\right)+\f{2}{\kappa r}\left(1-\f{b(r)}{r}\right)\right]\Phi^{\prime}(r)\nn&+&\f{2\lambda(r)b^\prime(r))}{r^2}-\f{b(r)}{\kappa r^3},\label{prexp}\\
p_t(r)&=&\f{1}{2\kappa r^3}\Bigg\{-4r^2(r-b(r))\left(\kappa\lambda(r)-\f{1}{2}\right)\left[\Phi^{\prime\prime}(r)+\Phi^{\prime2}(r)\right]+b^\prime(r)(4r\kappa\lambda(r)-r)+b(r)\nn&+&r\Phi^\prime(r)\left[rb^\prime(r)(2\kappa\lambda(r)-1)-2\kappa\lambda(r)(4r-3b(r))+2r-b(r)\right]\Bigg\}.\label{ptexp}
\eea
The above equations construct a system of four coupled differential equations to be solved for six unknowns $\{\rho(r),p_r(r),p_t(r),\Phi(r),b(r),\lambda(r)\}$. Hence, this system is under-determined and to solve it one needs to specify the functionality of two of the unknowns. We therefore proceed to deal with this task in the next sections and try to find physically reasonable wormholes solutions.
\subsection{Solutions with $\Phi(r)=0$}\label{phizero}
\begin{figure}
	\begin{center}
		\includegraphics[width=7.8cm]{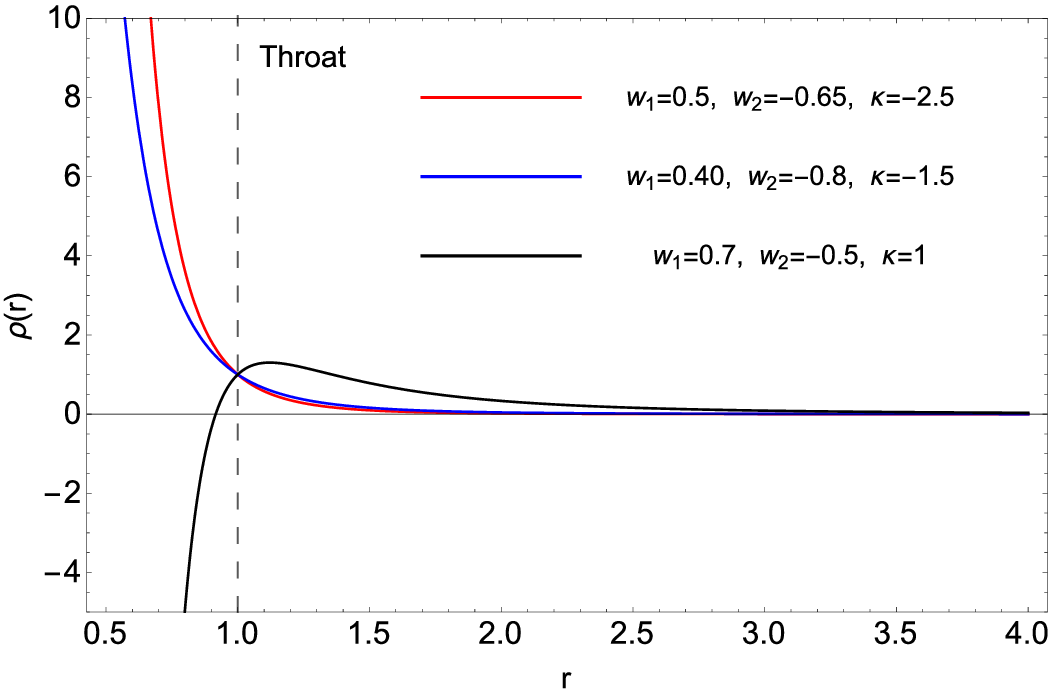}
		\caption{Plot of energy density for $\rho_0=r_0=1$. Right panel: The allowed region for the pair of parameters $(w_1,w_2)$ for $r_0=\kappa=1$, $\alpha=1$, $\beta=-0.3$ (blue region), $\beta=-0.2$ (red region) and $\beta=-0.1$ (gray region).}\label{fig1}
	\end{center}
\end{figure}

In this subsection we try to find a consistent solution for the system of differential equations, i.e., Eqs.~(\ref{consexp})-(\ref{ptexp}). From technical viewpoint building traversable wormhole models requires finding solutions to the field equations of the relevant gravitational theory. This can be done by defining specific set of boundary conditions along with some extra information in order to close the system of differential equations. A possible strategy to close the system is to determine the two metric functions that satisfy the required boundary conditions in order to obtain the energy-momentum distribution~\cite{mt}, see also~\cite{Lemos2003} and references therein. Other possibility for creating physically reasonable wormhole solutions is to provide an EoS and one of the metric functions. In this section we proceed in this manner and consider a vanishing redshift function along with linear equations of state for radial and tangential pressures as, $p_r(r)=w_1\rho(r)$ and $p_t(r)=w_2\rho(r)$ where $w_1$ and $w_2$ are constants. We therefore get three differential equations for three unknowns $\rho(r)$, $\lambda(r)$ and $b(r)$. Hence, the resulted three differential equations can be solved simultaneously with the solution given by
\bea\label{solphi0}
b(r)&=&r_0\left[\f{r}{r_0}\right]^\delta,~~~~~\rho(r)=\rho_1r^{\eta}+\rho_2\left(\f{r}{r_0}\right)^\gamma,~~~\delta=-\f{w_1+2w_2+3}{w_1-2w_2-1},~~~~~~\gamma=\f{4(w_2-w_1)}{w_1-2w_2-1},\nn
\lambda&=&\f{w_1+2w_2+1}{2\kappa(w_1+2w_2+3)},~~~~~~~\eta=\f{2(w_2-w_1)}{w_1},\nn\rho_1&=&\left[\rho_0r_0^{\f{2(w_1-w_2)}{w_1}}+\f{2r_0^{\f{-2w_2}{w_1}}}{\kappa(w_1-2w_2-1)}\right],~~~\rho_2=-\f{2}{\kappa(w_1-2w_2-1)r_0^2}
\eea
where $r_0$ is the radius of wormhole throat. The above solution contains two constants of integration that the first one can be found assuming the value of energy density at the throat as $\rho(r_0)=\rho_0$. The spherical surface $r=r_0$ has to satisfy the fundamental condition \ref{c1}, i.e., $b(r_0)=r_0$. In order to find the remained integration constant we have used this condition at the wormhole throat. The flare-out condition \ref{c2} at the throat leads to the inequality $\delta<1$ which also also satisfies the asymptotic flatness condition \ref{c4}. Also for $\delta<1$, the ratio $b(r)/r$ stays less than unity for $r>r_0$ hence Lorentzian signature is preserved, as required by condition \ref{c3}.
In the framework of classical GR, the fundamental flaring-out condition results in the violation of NEC. Such a violation can be surveyed by applying the focusing theorem on a congruence of null rays, defined by a null vector field $k^\mu$, where $k^\mu k_\mu=0$~\cite{khu,FLoboBook}. For the EMT presented in Eq.~(\ref{emtaniso}) the NEC is given by 
\bea
&&\rho(r)+p_{r}(r)\geq0,~~~~~\rho(r)+p_{t}(r)\geq0\label{nec}.
\eea
Also, for the sake of physical reliability of the solutions, we require that the wormhole configuration respects the WEC given by the following inequalities
\bea
&&\rho(r)\geq0,~~~~~~~~\rho(r)+p_{r}(r)\geq0,~~~~~~~~\rho(r)+p_{t}(r)\geq0.\label{wec}
\eea
We note that WEC implies the null form. The fulfillment of the second and third part of the above conditions requires that $w_1\geq-1$ and $w_2\geq-1$. Also the energy conditions at the throat take the form
\bea
\rho(r)\Big|_{r=r_0}&=&\rho_0\geq0,\label{energy0}\\
\rho(r)+p_{r}(r)\Big|_{r=r_0}&=&\rho_0(1+w_1)\geq0,\label{rhopluspr0}\\
\!\!\!\!\rho(r)+p_{t}(r)\Big|_{r=r_0}&=&\rho_0(1+w_2)\geq0\label{rhopluspt0}.
\eea
We note that the asymptotic flatness requires that no matter distribution is present at spatial infinity or $\rho(r\rightarrow\infty)=0$. This condition leads to $\eta<0$ and $\gamma<0$. In order that the energy density remains positive for $r>r_0$ we may consider three possible cases as follows:
\begin{enumerate}
\item $\eta<0~~~\land~~~\gamma<0~~~\land~~~\rho_1\geq0~~~\land~~~\rho_2\geq0$,\label{cond1}
\item $\eta<0~~~\land~~~\gamma<0~~~\land~~~|\eta|>|\gamma|~~~\land~~~\rho_1<0~~~\land~~~\rho_2\geq0~~~\land~~~\rho_1+\rho_2\geq0$,\label{cond2}
\item $\eta<0~~~\land~~~\gamma<0~~~\land~~~|\eta|<|\gamma|~~~\land~~~\rho_1\geq0~~~\land~~~\rho_2<0~~~\land~~~\rho_1+\rho_2\geq0$,\label{cond3}
\end{enumerate}	
Considering now the asymptotic flatness, energy conditions at the throat Eqs.~(\ref{energy0})-(\ref{rhopluspt0}), and flare-out condition we arrive at following inequalities for each of the above three conditions (we have set $r_0=1$):
\bea\label{confin1}
&&\rho_0>0\land \Bigg\{\left[\kappa \leq -\frac{2}{\rho_0}\land 0<w_1\leq 1\land -1\leq w_2\leq \frac{\kappa\rho_0(w_1-1)+2}{2 \kappa\rho_0}\right]\lor \Bigg[\frac{-2}{\rho_0}<\kappa <\frac{-1}{\rho_0}\land\nn&& \left[\left(w_1=\frac{-\kappa\rho_0-2}{\kappa\rho_0}\land w_2=-1\right)\lor \left(\frac{-\kappa\rho_0-2}{\kappa\rho_0}<w_1\leq 1\land -1\leq w_2\leq \frac{\kappa\rho_0(w_1-1)+2}{2 \kappa\rho_0}\right)\right]\Bigg]\lor\nn&&\left(\kappa =-\frac{1}{\rho_0}\land w_1=1\land w_2=-1\right)\Bigg\},
\eea
for condition \ref{cond1},
\bea\label{confin2}
&&\rho_0>0\land \Bigg\{\left[\kappa \leq-\frac{2}{\rho_0}\land 0<w_1<-\frac{1}{\kappa\rho_0}\land \frac{\kappa\rho_0(w_1-1)+2}{2 \kappa\rho_0}<w_2<-\frac{1}{2}(w_1+1)\right]\lor\nn &&\Bigg[-\frac{2}{\rho_0}<\kappa<-\frac{1}{\rho_0}\land \Bigg[\left(0<w_1<\frac{-\kappa\rho_0-2}{\kappa\rho_0}\land -1\leq w_2<-\frac{1}{2}(w_1+1)\right)\lor\nn &&\left(w_1=\frac{-\kappa\rho_0-2}{\kappa\rho_0}\land-1<w_2<-\frac{1}{2}(w_1+1)\right)\lor\Bigg(\frac{-\kappa\rho_0-2}{\kappa\rho_0}<w_1<-\frac{1}{\kappa\rho_0}\land\nn&& \frac{\kappa\rho_0(w_1-1)+2}{2\kappa\rho_0}<w_2<-\frac{1}{2}(w_1+1)\Bigg)\Bigg]\Bigg]\lor \left(-\frac{1}{\rho_0}\leq \kappa <0\land 0<w_1<1\land -1\leq w_2<-\frac{1}{2} (w_1+1)\right)\Bigg\},\nn
\eea
for condition \ref{cond2}, and
\bea\label{confin3}
\rho_0\geq 0~~\land~~\kappa >0~~\land~~0<w_1\leq 1~~\land~~-\frac{1}{2}(w_1+1)<w_2<\frac{w_1-1}{2},
\eea
for condition \ref{cond3}. Fig.~(\ref{fig1}) shows the behavior of energy density for different values of EoS parameters according to the above conditions. It is therefore seen that the supporting matter for wormhole configuration obeys the WEC (as long as $-1\leq w_1\leq 1$ and $-1\leq w_2\leq 1$), either at the throat and throughout the spacetime. A special type of solutions can be found for the case in which the matter distribution obeys the EoS of dark energy in tangential direction, i.e., $w_2=-1$. For this case $\delta=-1$ and the conditions \ref{cond1} and \ref{cond2} are reduced to the following inequalities, respectively
\bea\label{condsdark1}
&&\rho_0>0\land\nn&&\left[\left(\kappa \leq -\frac{2}{\rho_0}\land 0<w_1\leq 1\right)\lor \left(-\frac{2}{\rho_0}<\kappa <-\frac{1}{\rho_0}\land \frac{-\kappa\rho_0-2}{\kappa\rho_0}\leq w_1\leq 1\right)\lor \left(\kappa =-\frac{1}{\rho_0}\land w_1=1\right)\right],\nn
\eea
\bea\label{condsdark2}
\rho_0>0\land \left[\left(-\frac{2}{\rho_0}<\kappa <-\frac{1}{\rho_0}\land0<w_1<\frac{-\kappa\rho_0-2}{\kappa\rho_0}\right)\lor \left(-\frac{1}{\rho_0}\leq \kappa <0\land 0<w_1<1\right)\right],
\eea
The third condition does not provide physically reasonable values for the parameters $\{w_1,\rho_0,\kappa\}$. Hence, considering the above two conditions the spacetime metric Eq.~(\ref{evw}) is reduced to the following form
\be\label{mtmetr}
ds^2=-dt^2+\f{dr^2}{1-\f{r_0^2}{r^2}}+r^2d\Omega^2,
\ee
which represents the well-known Morris-Thorne spacetime~\cite{mt}. 
\subsection{Solutions with $\Phi(r)=\f{1}{2}\ln[\alpha+\beta\left(\f{r_0}{r}\right)]$}\label{phinzero}
Let us now investigate wormhole solutions with a non-vanishing redshift function. Substituting then for $\Phi(r)$ into Eqs.~(\ref{consexp})-(\ref{ptexp}) together with considering linear EoS in radial and tangential directions we arrive at a system of coupled differential equations for the three unknowns $\rho(r)$, $\lambda(r)$ and $b(r)$. This system can be solved analytically and the solution is found as
\bea\label{gensolb}
\rho(r)&=&{\rm C}_1r^{\f{4w_2-3w_1+1}{2w_1}}(\alpha r+\beta r_0)^{-\f{w_1+1}{2w_1}}\nn&+&\f{{\rm C}_2(\alpha r+\beta r_0)}{r^2}\left[2r\alpha(w_1-2w_2-1)+r_0\beta(3w_1-4w_2-1)\right]^{\f{2w_2-3w_1-1}{w_1-2w_2-1}},\label{rhosol}\\
\lambda(r)&=&\f{2r\alpha(w_1+2w_2+1)+r_0\beta(w_1+4w_2+1)}{4\kappa\left[r\alpha(w_1+2w_2+3)+2r_0\beta(w_2+1)\right]},\label{lambdasol}\\
b(r)&=&\Bigg\{\f{\kappa r^3}{4}(\alpha r+\beta r_0)w_1\left[r\alpha(w_1-2w_2-1)+\f{3}{2}\left(w_1-\f{4}{3}w_2-\f{1}{3}\right)\beta r_0\right]^2\rho^\prime(r)\nn&+&\f{\kappa r^2}{2}\left[\alpha(w_1-w_2)r+\f{3}{4}\left(w_1-\f{4}{3}w_2-\f{1}{3}\right)r_0\beta\right]\left[r\alpha(w_1-2w_2-1)+\f{3}{2}\left(w_1-\f{4}{3}w_2-\f{1}{3}\right)r_0\beta\right]^2\rho(r)\nn&-&r_0\beta\Big[\alpha^2(w_1+2w_2+1)(w_1-w_2)r^2+\left[w_1^2+w_1\left(\f{3}{4}+2w_2\right)-4\left(w_2+\f{1}{4}\right)^2\right]\alpha\beta rr_0\nn&+&\f{3}{8}(w_1+4w_2+1)\beta^2r_0^2\left(w_1-\f{4}{3}w_2-\f{1}{3}\right)\Big]\Bigg\}/\Bigg[\alpha\Big(\alpha^2(w_1+2w_2+1)(w_1-w_2)r^2\nn&+&\alpha\beta rr_0\left[w_1^2+w_1\left(2w_2+\f{3}{4}\right)-4\left(w_2+\f{1}{4}\right)^2\right]+\f{3}{8}(w_1+4w_2+1)\beta^2r_0^2\left(w_1-\f{4}{3}w_2-\f{1}{3}\right)\Bigg].\label{bsol}
\eea
where ${\rm C}_1$ and ${\rm C}_2$ are integration constants. We find the constant ${\rm C}_1$ using the value of energy density at the throat, i.e., $\rho(r_0)=\rho_0$ and the constant ${\rm C}_2$, using the condition $b(r_0)=r_0$. A straightforward but lengthy calculation then gives
\bea
b(r)&=&\left(r+\f{\beta}{\alpha}r_0\right)\left[\f{A(r)}{A(r_0)}\right]^{1-\epsilon_1}-\f{\beta}{\alpha}r_0,\label{bsolfinal}\\
\rho(r)&=&\f{1}{\kappa A(r_0)r^2}\Bigg\{-4(\alpha r+\beta r_0)\left[\f{A(r_0)}{A(r)}\right]^{\epsilon_1}+\rho_3\left(\f{r}{r_0}\right)^{\epsilon_2}\left[\f{(\alpha+\beta)r_0}{\alpha r+\beta r_0}\right]^{\epsilon_3}\Bigg\},\label{rhofinal}
\eea
where
\bea\label{AC}
\epsilon_1&=&\f{3w_1-2w_2+1}{w_1-2w_2-1},~~~~~\epsilon_2=\f{4w_2+w_1+1}{2w_1},~~~~~\epsilon_3=\f{w_1+1}{2w_1},\nn\rho_3&=&4r_0(\alpha+\beta)+\kappa r_0^2\rho_0A(r_0),~~~~~~~~~A(r)=2r\alpha(w_1-2w_2-1)+r_0\beta\left(3w_1-4w_2-1\right).
\eea
It can be easily checked that $b(r_0)=r_0$ and $\rho(r_0)=\rho_0$. Considering the above solution, the following points can be argued:
\begin{enumerate}[label=\Roman*]
	\item. From Eq.~(\ref{bsolfinal}) the flare-out condition leads to the following inequality at the throat\label{p1}
	\bea\label{flarecond}
	-\f{2\alpha(2w_2+w_1+3)+\beta(4w_2+w_1+5)}{2\alpha(w_1-2w_2-1)+\beta(3w_1-4w_2-1)}<1.
	\eea
	\item. From Eqs.~(\ref{bsolfinal}) and (\ref{rhofinal}) the asymptotic flatness of the solution implies\label{p2}
	\be\label{epsis}
	\epsilon_1>1~~~~\land~~~~\epsilon_2<0~~~~\land~~~~\epsilon_3>0.
	\ee
	\item. For the temporal component of metric at spatial infinity we have\label{p3}
	\be\label{tempmet}
	\lim_{r\rightarrow\infty}{\rm e}^{2\Phi(r)}=\lim_{r\rightarrow\infty}\left(\alpha+\beta\f{r_0}{r}\right)=\alpha>0.
	\ee
	\item. The WEC at wormhole throat implies\label{p4}
	\be\label{wec0}
     \rho_0\geq0,~~~~~~~~~-1\leq w_1\leq 1,~~~~~~~~~-1\leq w_2\leq 1,
	\ee
	\item. In order that the energy density remains positive for $r>r_0$, we assume that $\rho_3>-4r_0(\alpha+\beta)$ and note that the first term in curly brackets of Eq.~(\ref{rhofinal}) decays as $1/r^{\epsilon_1-1}$ and the second term as $1/r^{\epsilon_3+|\epsilon_2|}$. Now we may consider the condition $\epsilon_3+|\epsilon_2|<\epsilon_1-1$ so that if at the throat $\rho(r_0)=\rho_0>0$, it then remains positive throughout the spacetime.\label{p5}
\end{enumerate}

\par
\begin{figure}
	\begin{center}
		\includegraphics[width=7cm]{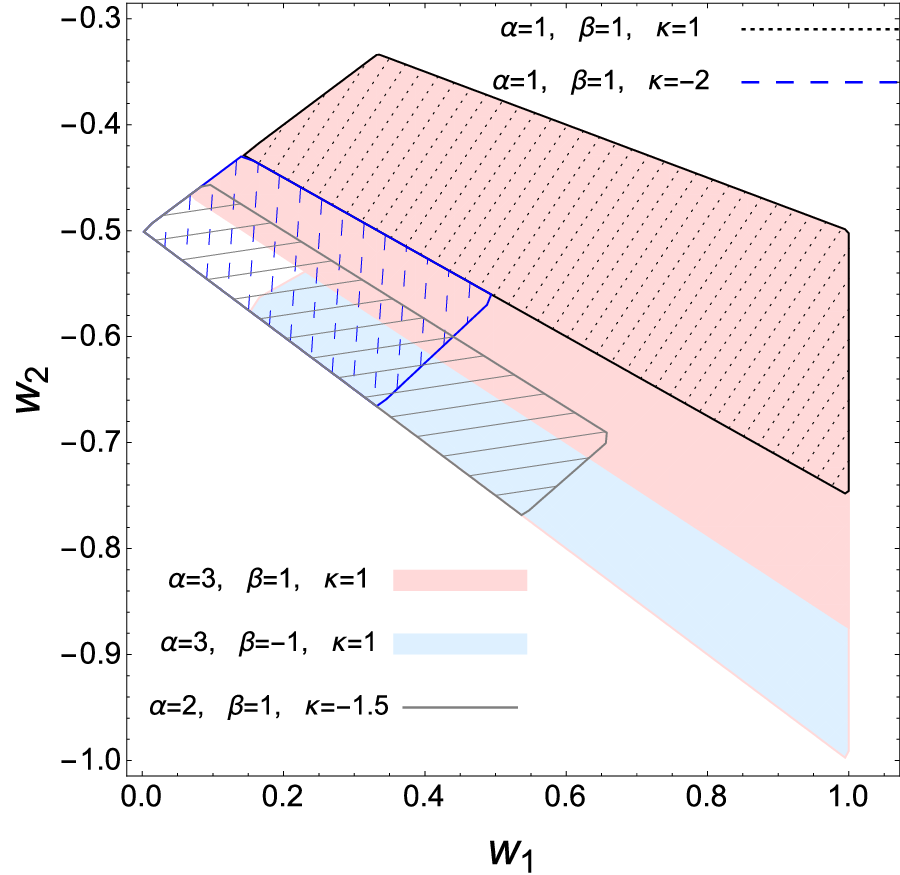}
		\includegraphics[width=7.6cm]{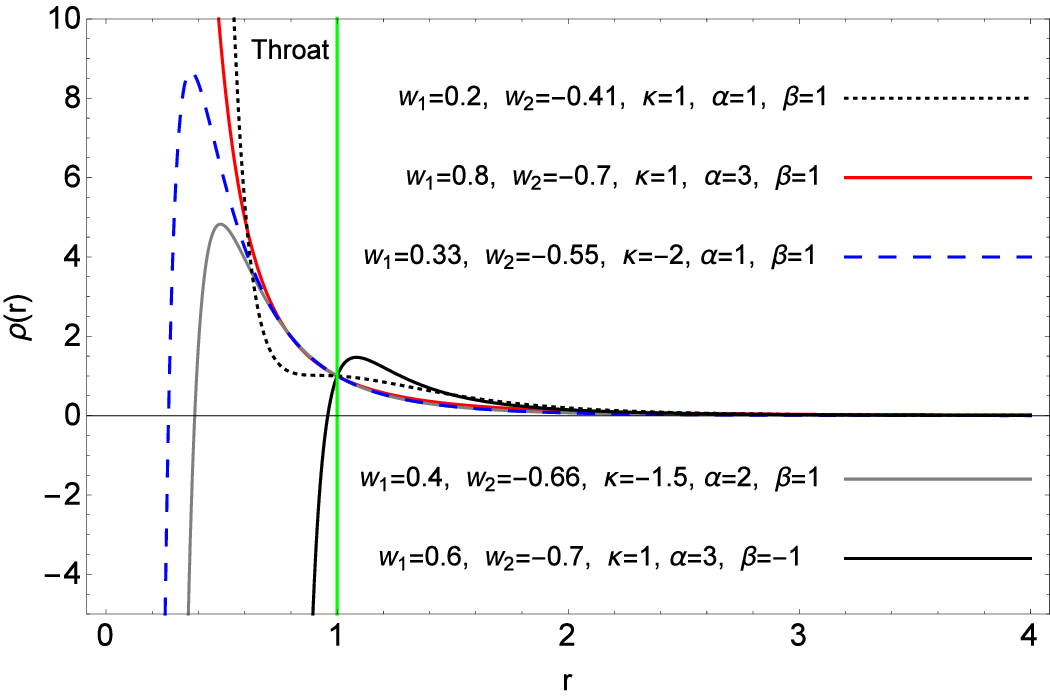}
		\caption{Left panel: The allowed values of EoS parameters for $r_0=\rho_0=1$. Right panel: Behavior of energy density for the allowed values of model parameters chosen from the left panel.}\label{fig2}
	\end{center}
\end{figure}
\begin{figure}
	\begin{center}
		\includegraphics[width=7.6cm]{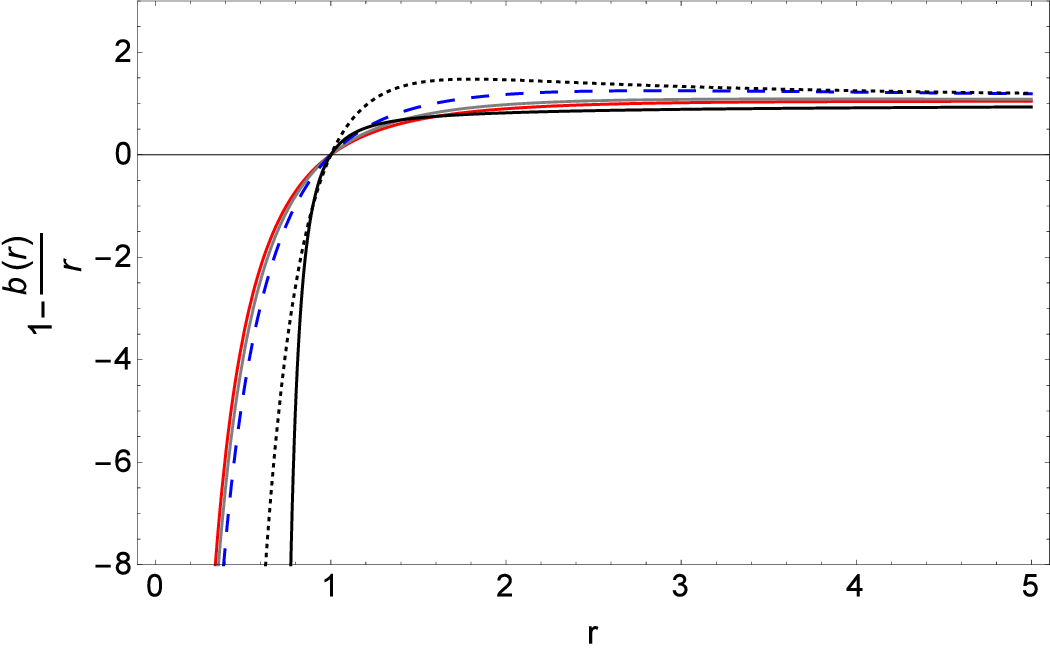}
		\includegraphics[width=7.6cm]{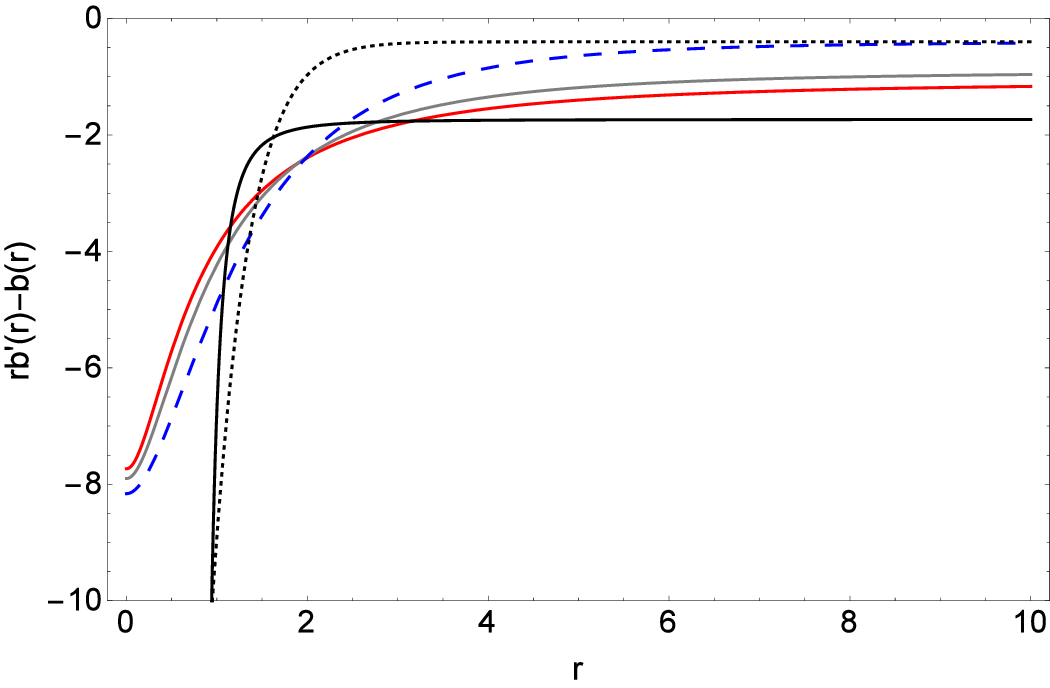}
		\caption{The behavior of inverse of the radial metric component (left panel) and the flare-out condition (right panel) for the same values of model parameters as of the right panel in Fig.~(\ref{fig2}).}\label{fig3}
	\end{center}
\end{figure}
\begin{figure}
	\begin{center}
		\includegraphics[width=7.6cm]{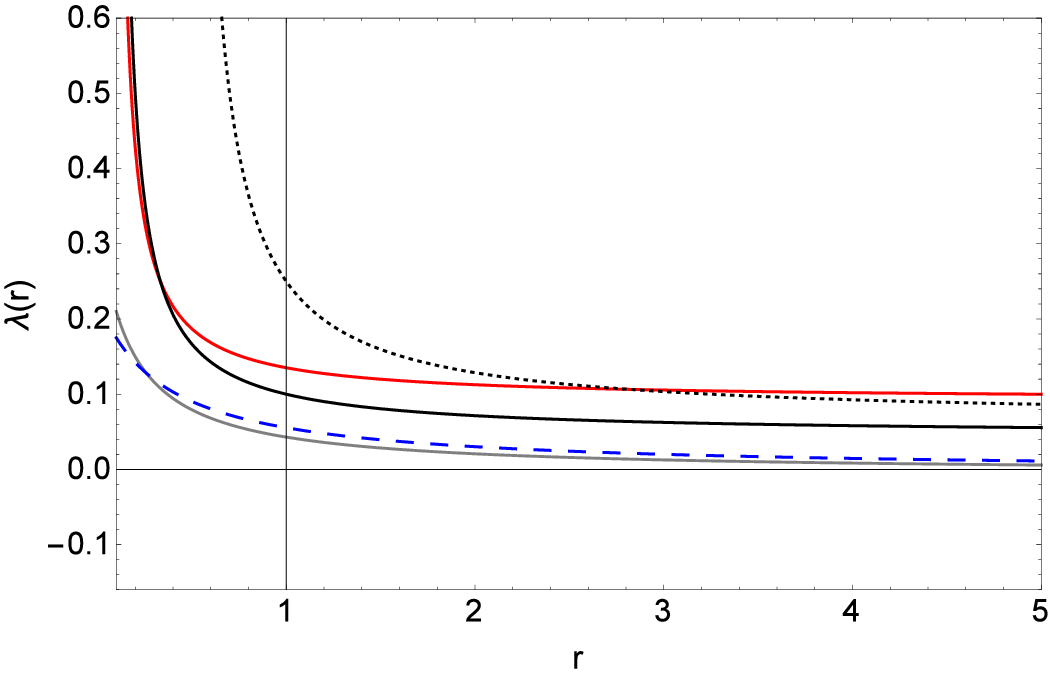}
		\caption{The behavior of Rastall parameter for the same values of model parameters as of the right panel in Fig.~(\ref{fig2}).}\label{fig33}
	\end{center}
\end{figure}

The left panel in Fig.~(\ref{fig2}) presents the allowed regions of the pair $(w_1,w_2)$ according to the conditions (\ref{p1}-\ref{p5}) stated above. As we observe, wormhole configurations with a non-vanishing redshift function satisfy the WEC in both directions. The right panel shows the behavior of energy density as a function of radial coordinate for different values of model parameters subject to the conditions (\ref{p1}-\ref{p5}). It is therefore seen that considering the condition \ref{p4}, the WEC and NEC are satisfied at the throat and throughout the spacetime and thus, there is no need of introducing exotic matter in order to construct the present wormhole solutions. The left panel in Fig.~(\ref{fig3}) shows the behavior of radial metric component, i.e., $g_{rr}=\left(1-b(r)/r\right)^{-1}$. We observe that $g_{rr}^{-1}$ is positive for $r>r_0$ and thus the metric signature is preserved for radii bigger than the throat radius. We note that the metric is asymptotically flat since, as $r\rightarrow\infty$, both the temporal and radial components of the metric tend to unity. To be a solution of a wormhole, one needs to impose that the throat flares out. This can be seen in the right panel of Fig.~(\ref{fig3}) as the satisfaction of flare-out condition. In Fig.~(\ref{fig33}), we have plotted the behavior of Rastall parameter where we observe that this parameter decreases monotonically as one moves away from the wormhole configuration. This behavior may be interpreted in such a way that the effects of a running mutual interaction between matter and geometry, encoded in the variable Rastall coupling parameter, increase in the limit of approach to the wormhole throat and these effects can contribute to constructing wormhole configurations without the need of exotic matter. Moreover, asymptotically the Rastall parameter tends to a constant that the value of which is decided by the EoS parameters of the matter content of the wormhole configuration, see the solution Eq.~(\ref{solphi0}) with vanishing redshift function. We therefore conclude that for the present class of wormhole solutions, the satisfaction of WEC and NEC for a non-vanishing redshift function requires a varying coupling between matter and geometry.    
\section{Timelike and Null Geodesics}\label{glens}
\begin{figure}
	\begin{center}
		\includegraphics[width=7.7cm]{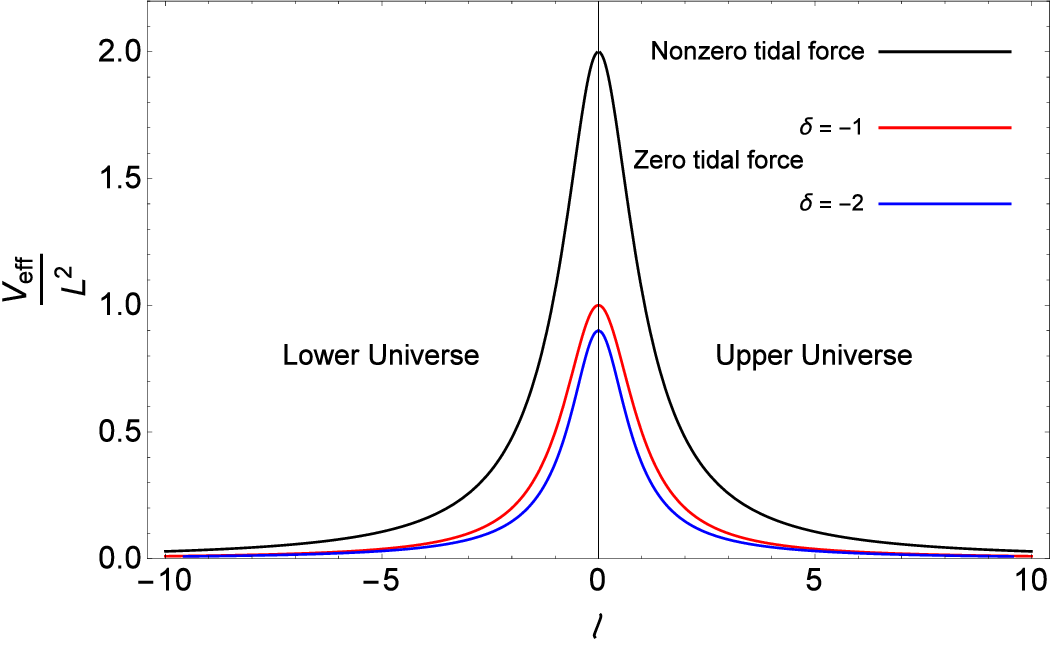}
		\includegraphics[width=7.7cm]{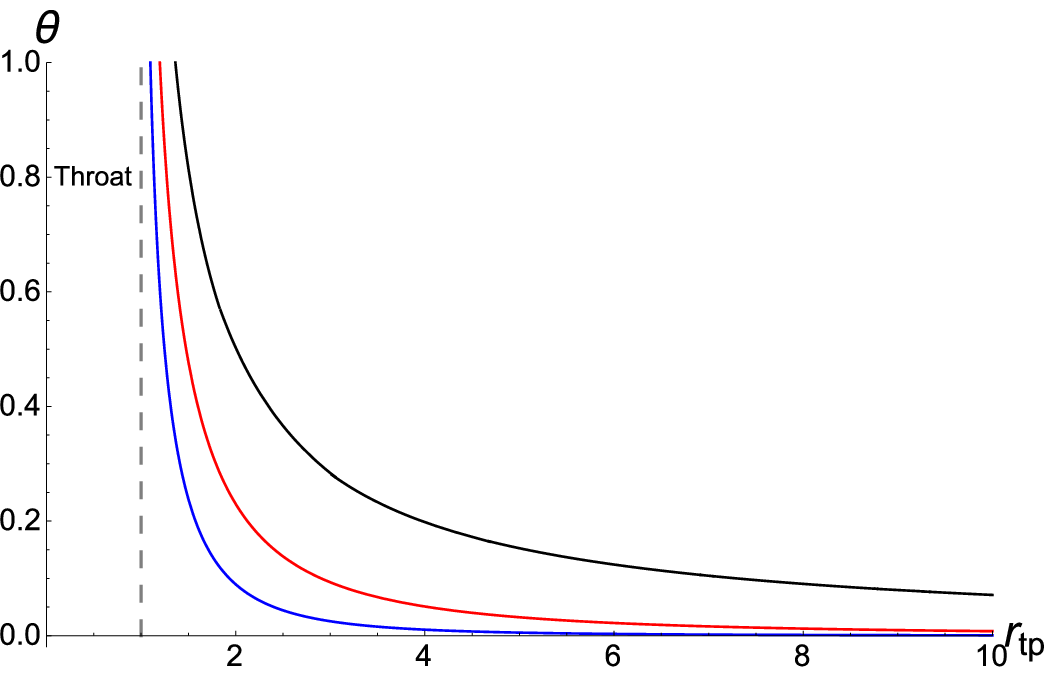}
		\caption{Left panel: The behavior of effective potential for null geodesics against proper radial distance for both zero and nonzero tidal force solutions. For the black curve we have set $\epsilon_1=3$, $w_2=-1$, $\alpha=3$, $\beta=-1$, $\kappa=1$, $r_0=1$ and the EoS in radial direction obeys the light blue region in the left panel of Fig.~(\ref{fig2}). Right panel: Behavior of deflection angle against radial coordinate for the same values of model parameters as of the left panel.}\label{fig44}
	\end{center}
\end{figure}

Now, we shall examine the behavior of timelike and null geodesics in the wormhole spacetime. To this aim we begin with the Lagrangian ${\mathcal L}=\f{1}{2}g_{\mu\nu}\dot{x}^\mu\dot{x}^\nu$ associated to the metric (\ref{evw}),
\be\label{Lagran}
2{\mathcal L}=g_{\mu\nu}\dot{x}^\mu\dot{x}^\nu=-{\rm e}^{2\Phi(r)}\dot{t}^2+\left(1-\f{b(r)}{r}\right)^{-1}\dot{r}^2+r^2\dot{\phi}^2,
\ee
where an overdot denotes derivative with respect to the curve parameter $\zeta$. Because of the spherical symmetry we consider the equatorial plane $\theta=\pi/2$. The Lagrangian ${\mathcal L}(\dot{x},x)$ is constant along a geodesic curve, therefore one can set ${\mathcal L}(\dot{x},x)=\xi$ so that the paths in spacetime can be classified to timelike ($\xi=-1$), null ($\xi=0$) and spacelike ($\xi=1$) geodesics~\cite{Hobs}. Since the Lagrangian (\ref{Lagran}) does not depend on the variables $t$ and $\phi$, we have two constants of motion given by
\bea\label{ELcons}
\f{\partial{\mathcal L}}{\partial\dot{t}}=-{\rm e}^{2\Phi(r)}\dot{t}=-E,~~~~~~~~~~~~\f{\partial{\mathcal L}}{\partial\dot{\phi}}=r^{2}\dot{\phi}=L,
\eea
where $E$ and $L$ are the energy and angular momentum of the particle moving on its orbit, respectively. Null geodesics are of great importance as they can reveal the interesting features of a compact object in the Universe, i.e., its gravitational lensing effects. Wormholes are no exception as lensing effects of these objects can provide a way to search for their observational signatures and more interestingly, distinguish them from black holes~\cite{wormbhtsu}. Substituting for $(\dot{t},\dot{\phi})$ from Eq.~(\ref{ELcons}) into Eq.~(\ref{Lagran}) we arrive at the following equation for particle trajectory 
\be\label{effequation}
\dot{r}^2={\rm e}^{-2\Phi(r)}\left[1-\f{b(r)}{r}\right]\left[E^2-V_{\rm eff}(r)\right]=0,~~~~~~~~~~~V_{\rm eff}(r)={\rm e}^{2\Phi(r)}\left[\f{L^2}{r^2}-\xi\right],
\ee
where $V_{\rm eff}$ is the effective potential. To investigate lensing features of the obtained wormhole solutions we consider the condition on null geodesics ($\xi=0$) and use the second part of Eq.~(\ref{ELcons}) to get
\be\label{drdphi}
\left(\f{dr}{d\phi}\right)^2=r^2\left[1-\f{b(r)}{r}\right]\left[\f{r^2}{r_{\rm im}^2}{\rm e}^{-2\Phi(r)}-1\right],
\ee
where $r_{\rm im}=L/E$ is the impact parameter. For a photon coming from a distant source, taking a turn at $r_{\rm tp}$ and escaping then to a faraway observer the deflection angle is given by~\cite{SWeinbergbook}
\be\label{defangleequa}
\theta(r_{\rm tp})=2\int_{r_{\rm tp}}^{\infty}\f{{\rm e}^{\Phi(r)} dr}{\left[\left(r^2-rb(r)\right)\left(\f{r^2}{r_{\rm im}^2}-{\rm e}^{2\Phi(r)}\right)\right]^{\f{1}{2}}}-\pi,
\ee 
where the turning point $r_{\rm tp}$ is given by the condition $dr/d\phi=0$. This condition leads to the following relation between impact parameter and turning point 
\be\label{rtpimp}
r_{\rm im}=r_{\rm tp}{\rm e}^{-\Phi(r_{\rm tp})}.
\ee
Strong gravitational lensing occurs when $r_{\rm tp}$ is close to the location of the photon sphere, i.e., a radius at which
light can bend in angles excess of $2\pi$~\cite{Bozza}. The conditions for circular photon orbits are given by $\dot{r}=0$ and $\ddot{r}=0$ and stable (unstable) photon orbits satisfy the additional condition $\dddot{r}<0(>0)$~\cite{Hasse-Perlick,Perlicklvr}. These conditions can be reexpressed in terms of effective potential as
\be\label{photoneff}
V_{\rm eff}(r)\Big|_{r=r_{\rm ph}}=E^2,~~~V^\prime_{\rm eff}(r)\Big|_{r=r_{\rm ph}}=0,~~~~V^{\prime\prime}_{\rm eff}(r)\Big|_{r=r_{\rm ph}}<0.
\ee
where $r_{ph}$ denotes the radius of photon sphere. Therefore, the local maximum points of the effective potential provide unstable photon orbits, while those of local minima give stable ones. The family of unstable photon orbits construct the photon sphere. From Eq.~(\ref{effequation}) we find that the condition $\dot{r}=0$ is also satisfied at wormhole throat since $b(r_0)=r_0$. Hence, one may deduce that the throat can play the role of an effective photon sphere when the conditions $\ddot{r}=0$ and $\dddot{r}>0$ are fulfilled. In~\cite{Shaikh-novel}, it has been shown that this happens when the following conditions on effective potential are satisfied
\be\label{photoneffsh}
V_{\rm eff}(r)\Big|_{r=r_0}=E^2,~~~~~~~~V^\prime_{\rm eff}(r)\Big|_{r=r_0}<0,
\ee
where use has been made of the flare-out condition $b^\prime(r_0)<1$. If the photon sphere is situated at the throat then $V_{\rm eff}$ must admit a maximum value at $r=r_0$. To better understand the issue one may switch to radial proper distance given in Eq.~(\ref{radpropdis}). Hence, Eq.~(\ref{effequation}) reads
\be\label{eqeffl}
\dot{\ell}^2{\rm e}^{2\Phi(r(\ell))}+V_{\rm eff}=E^2,~~~~~~~~~~~V_{\rm eff}=\f{L^2{\rm e}^{2\Phi(r(\ell))}}{r^2(\ell)}.
\ee
Therefore at the throat, we get the following conditions
\bea\label{cond9}
&&V_{\rm eff}(r)\Big|_{r=r_0}=E^2,~~~~~~~~\f{dV_{\rm eff}}{d\ell}\Big|_{\ell=0}=\pm\left[1-\f{b(r)}{r}\right]^{\f{1}{2}}V^\prime_{\rm eff}(r)\Big|_{r=r_0}=0,\nn
&&\f{d^2V_{\rm eff}}{d\ell^2}=\left[1-\f{b(r)}{r}\right]^{\f{1}{2}}V^{\prime\prime}_{\rm eff}(r)\Big|_{r=r_0}\!\!\!\!+\f{1}{2}\left[\f{b(r)}{r^2}-\f{b^\prime(r)}{r}\right]V^\prime_{\rm eff}(r)\Big|_{r=r_0}\!\!\!\!=\f{1-b^\prime(r_0)}{2r_0}V^\prime_{\rm eff}(r)\Big|_{r=r_0}<0,\nn
\eea
where we have considered the conditions $b^\prime(r_0)<1$ and $b(r_0)=r_0$. Now, we proceed to find the behavior of effective potential as a function of radial proper distance. To this aim we have to get the proper distance in terms of radial coordinate through Eq.~(\ref{radpropdis}). For zero tidal force solutions Eq.~(\ref{solphi0}), we have 
\bea\label{ellzero}
\ell(r)=\mp\f{2r\sqrt{1-r^{1-\delta}}}{(\delta-3)\sqrt{1-r^{\delta-1}}}{_2}F_1\left[\f{1}{2},\f{\delta-3}{2(\delta-1)},\f{5-3\delta}{2-2\delta},r^{1-\delta}\right]\pm\sqrt{\pi}\f{\Gamma\left[\f{4-3\delta}{2(1-\delta)}\right]}{\Gamma\left[\f{2-\delta}{1-\delta}\right]},
\eea  
where we have set $r_0=1$ for simplicity. For $\delta=-1$ which corresponds to Morris-Thorne wormhole~\cite{mt} the integration can be easily done with the solution $\ell(r)=\pm\sqrt{r^2-1}$. We note that for this case the EoS in tangential direction is that of dark energy and the EoS in radial direction has to obey the conditions given in Eqs.~(\ref{condsdark1}) and (\ref{condsdark2}). Then the effective potential is obtained as: $V_{\rm eff}(\ell)=L^2/(\ell^2+1)$. For nonzero redshift solutions Eq.~(\ref{bsolfinal}), the integral (\ref{radpropdis}) cannot be solved analytically in terms of elementary functions. Hence, we proceed to consider special cases, for example, the case with $\epsilon_1=3$ which leads to $w_2=-1$ and $w_1$ has to obey the allowed regions given in the left panel of Fig.~(\ref{fig2}). For this case we choose the model parameters related to the light blue region of this figure, i.e., $\alpha=3$, $\beta=-1$ and $\kappa=1$. Then, the integration can be performed with the following result
\be\label{nonzeroell}
\ell(r)=\pm\f{3-12r+9r^2+\sqrt{3}\sqrt{r-1}\sqrt{3r-1}\arcsinh\!\left[\sqrt{\f{3}{2}}\sqrt{r-1}\right]}{3\sqrt{3}(2r-1)\sqrt{\f{3r^2-4r+1}{(1-2r)^2}}}.
\ee
In the left panel of Fig.~(\ref{fig44}) we have plotted the effective potential in units of angular momentum squared versus radial proper distance. It is seen that the effective potential admits a maximum at the throat i.e., $\ell=0$, hence, the throat acts as a photon sphere. The right panel shows the behavior of deflection angle as a function of turning point coordinate radius. We observe that the more the turning point decreases, the more the deflection angle grows. As the turning point tends to the throat radius, the deflection angle increases and diverges at the wormhole throat where an unstable photon sphere is present. In such a situation, the wormhole configuration can produce infinite number of relativistic images of an appropriately placed source of light. This infinite sequence corresponds to infinitely many light rays whose limit curve asymptotically reaches the unstable photon sphere~\cite{Hasse-Perlick}. In this situation, since the photon sphere coincides with the wormhole throat, such a sphere can be detected utilizing thoroughly and carefully designed modern instruments~\cite{Perlicklvr},\cite{highsensinstru}, providing thus, possible observational proofs for the existence of the wormhole.
\par
\begin{figure}
	\begin{center}
		\includegraphics[width=7.7cm]{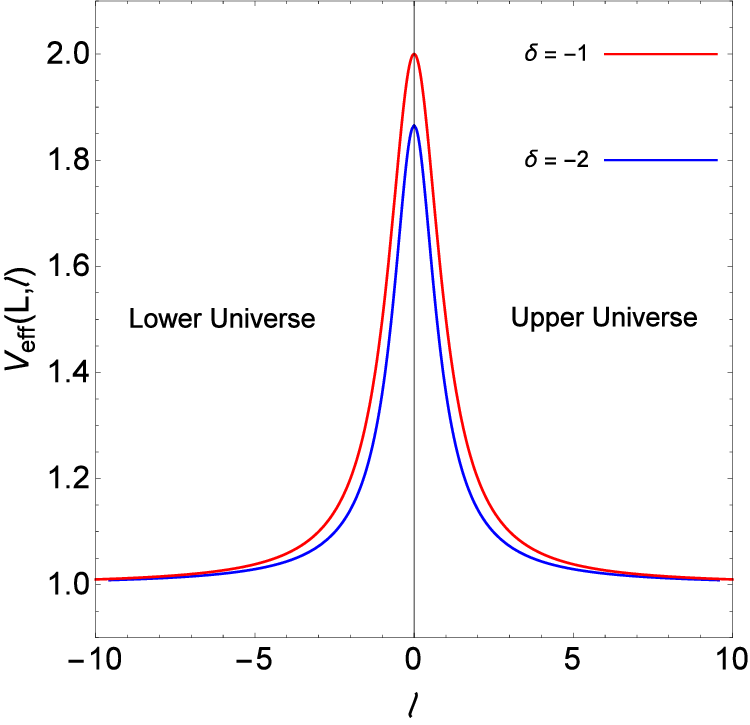}
		\includegraphics[width=7.7cm]{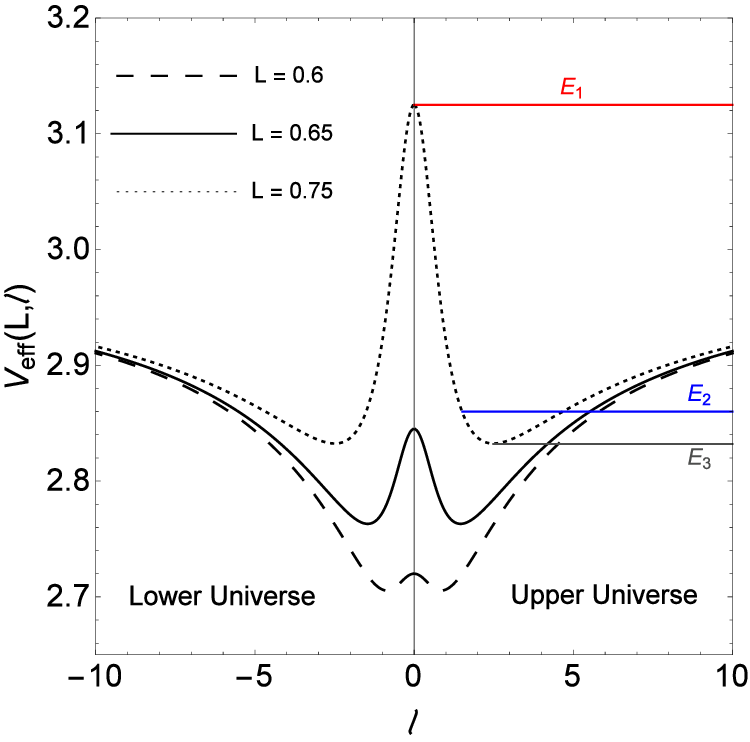}
		\caption{Left panel: The behavior of effective potential for timelike geodesics against proper radial distance for zero tidal force solutions. We have set $L=1$. Right panel: Behavior of effective potential for timelike geodesics for nonzero tidal force solutions. Numerical values for local minima are given as, $\ell=0.8545$ for $r_\star=1.3739$ and $L=0.6$, $\ell=1.4584$ for $r_\star=1.8497$ and $L=0.65$ and $\ell=2.4923$ for $r_\star=2.7646$ and $L=0.75$. Also, using Eq.~(\ref{veff2ndell1}) the values of second derivative of the effective potential in terms of proper distance for the corresponding values of $(r_\star,L)$ are found as: $d^2V_{\rm eff}/d\ell^2=0.084, 0.070, 0.030$.}\label{fig55}
	\end{center}
\end{figure}

Finally, for timelike geodesics we consider the case with $\xi=-1$ for which the equation governing orbital motion of particles i.e., Eq.~(\ref{eqeffl}) reads
\be\label{effpotime}
\left(\f{d\ell}{d\phi}\right)^2=\f{r(\ell)^4{\rm e}^{-2\Phi(r(\ell))}}{L^2}\left[E^2-V_{\rm eff}(L,\ell)\right],~~~~~~~~V_{\rm eff}(L,\ell)={\rm e}^{2\Phi(r(\ell))}\left[\f{L^2}{r^2(\ell)}+1\right].
\ee
Indeed, the above radial geodesic equation can be interpreted as a classical scattering problem with a potential barrier that depends on angular momentum. Hence, in traversable wormhole spacetimes, particles can pass through the wormhole throat from one region of the Universe to other one. This can occur for a particle moving on timelike geodesic if, $E^2>V_{\rm eff}(L,0)$. Also, for those geodesics reflected back by the potential barrier at the throat, we have $E^2<V_{\rm eff}(L,0)$. One then may recognize that the proper radius of closest approach is given by the turning point $\ell=\ell_{\rm tp}$ at which $dr/d\phi=0$. This gives the condition $E^2=V_{\rm eff}(L,\ell_{\rm tp})$. For zero tidal force solutions and the case in which $\delta=-1$, we get the effective potential as
\be\label{effpotdm1}
V_{\rm eff}(L,\ell)=\f{L^2+\ell^2+1}{\ell^2+1},
\ee
whereby we get
\be\label{deffpotdm1}
\f{dV_{\rm eff}(L,\ell)}{d\ell}\Big|_{\ell=0}=0,~~~~~~~~~~~\f{d^2V_{\rm eff}(L,\ell)}{d\ell^2}\Big|_{\ell=0}=-2L^2<0.
\ee
From the first part of the above expression we find that the effective potential admits an extremum at the throat. The second part shows that this potential possesses a global maximum at the throat which corresponds to unstable orbits. The left panel in Fig.~(\ref{fig55}) displays the behavior of effective potential versus proper radial distance, where we observe that, $\ell=0$ is the only turning point in the potential diagram. This point separates those geodesics approaching the wormhole from spatial infinity $\ell=\infty$, in the upper Universe, and are reflected back to $\ell=\infty$ in the very Universe, from those that pass through the wormhole throat and reach the spatial infinity of the lower Universe $\ell=-\infty$. Such a situation can also occur for zero tidal force solutions with $\delta=-2$, see the blue curve. For the solutions with nonzero tidal force we use the same parameters as the left panel of Fig.~(\ref{fig44}), the effective potential then reads $V_{\rm eff}(L,\ell)=(3r(\ell)-1)(L^2+r(\ell)^2)/r(\ell)^3$. This potential admits a local minimum at $r_{\star}=3L^2+\sqrt{3}L\sqrt{3L^2-1}$ where $V^{\prime\prime}_{\rm eff}(L,r_\star)>0$ and this local minimum is outside of the throat, i.e., $r_\star>1$ for $L>1/\sqrt{3}$, see the right panel in Fig.~(\ref{fig55}). It can be shown that the concavity of the effective potential is upward, i.e.,
\be\label{veff2ndell1}
\f{d^2V_{\rm eff}(L,\ell)}{d\ell^2}\Bigg|_{r=r_\star}=V^{\prime\prime}_{\rm eff}\left(1-\f{b}{r}\right)\Bigg|_{r=r_\star}+\f{V^\prime_{\rm eff}}{2r^2}(b-rb^\prime)\Bigg|_{r=r_\star}=V^{\prime\prime}_{\rm eff}\left[1-\f{b(r_\star)}{r_\star}\right]>0.
\ee
Also, the more the angular momentum increases the closer the local minimum to the throat. At the throat potential achieves a maximum where the first part of Eq.~(\ref{deffpotdm1}) holds and the second derivative of the effective potential with respect to the proper distance is negative. This can be checked through the following calculations, noting that $dr=\sqrt{(1-b/r)}d\ell$,
\be\label{veff2ndell}
\f{d^2V_{\rm eff}(L,\ell)}{d\ell^2}\Bigg|_{r=r_0=1}=V^{\prime\prime}_{\rm eff}\left(1-\f{b}{r}\right)\Bigg|_{r=r_0=1}+\f{V^\prime_{\rm eff}}{2r^2}(b-rb^\prime)\Bigg|_{r=r_0=1}=\f{V^\prime_{\rm eff}}{2}(1-b^\prime(r_0))<0,
\ee
where use has been made of the conditions \ref{c1} and \ref{c2}. Regarding the right panel of Fig.~(\ref{fig55}) we may have the following scenarios for particle orbits: A particle with total energy $E>E_1$ coming from spatial infinity $\ell=\infty$ can go through the wormhole throat and continue its trip to the spatial infinity of the other Universe, i.e., $\ell=-\infty$. The orbit of particles with $E=E_1$ and $L=3/4$ are subject to the unstable circular orbit at the throat and a perturbation can make the particle to escape to each of the Universes. This can also occur for the other values of angular momentum, i.e., particles with total energy $E=V_{\rm eff}(0.6,0)$ (dashed curve) and $E=V_{\rm eff}(0.65,0)$ (solid curve). Orbits of particles with $E=E_2$ correspond to oscillatory bound orbits and those with $E=E_3$ correspond to circular bound orbits. In the former case, the particle moves on an orbit that oscillates around the radius of a stable circular orbit at the minimum of the potential. For these orbits the particle can never cross the wormhole throat due to the potential barrier and thus, it is subject to stay in the same Universe. 
\section{Concluding Remarks}\label{concluding}
In the framework of GR, the satisfaction of flare-out condition requires the usage of exotic matter which violates the standard energy conditions as respected by normal matter. However, it has been shown that modified gravity theories provide a suitable framework for construction of traversable wormholes without the need of exotic matter or at least minimal usage of such type of matter, see e.g.,~\cite{ectwormhole1,ectwormhole},~\cite{lobocgqreview} and~\cite{wormmodif}. Motivated by these developments, in the present work, we tried to build and study static spherically symmetric wormhole configurations in GRG. We then tried to find exact spacetimes that admit suitable wormhole configurations, considering the field equations of GRG Eqs.~(\ref{CoVDiv}) and (\ref{genrasfieldeqs}). Two classes of exact solutions were obtained and discussed. In the first one, assuming a zero redshift function along with linear EoS between pressure profiles and energy density, we found exact zero tidal force wormhole solutions. We then provided the allowed values of EoS parameters for which the energy conditions as well as conditions on physical reasonability of wormhole solutions are met. The second class of our solutions deals with wormhole structures with nonzero tidal force. For these type of solutions we found that the supporting matter for wormhole geometry satisfies the WEC and NEC, provided that the EoS parameters in radial and tangential directions respect the allowed regions as provided by the left panel of Fig.~(\ref{fig2}). The study of wormholes in Rastall gravity has been performed by some authors for example, wormhole solutions for traceless fluid in the Rastall background~\cite{wormrastrace}, and wormhole solutions in original version of Rastall gravity~\cite{morad1,morad2}. These works present examples of wormholes that can exist in Rastall gravity with matter satisfying or violating NEC or even negative energy density at the throat~\cite{morad1}. However, the solutions obtained in the present article respect the WEC and NEC at the throat and throughout the spacetime. This result may be interpreted as the consequences of mutual dynamic interaction between matter and geometry which is encoded in the variable Rastall parameter. Such a dynamic interaction can also provide interesting cosmological as well as astrophysical models, e.g., successful description of the late time accelerated expansion of the Universe~\cite{morad3,moraddyn,Lin2020}, evolution of cosmological structures~\cite{struformgrg}, regular cosmological model in the context of Einstein static Universe~\cite{shaban}, nonsingular collapse scenario~\cite{ZiaRas} and regular black hole solutions~\cite{regbhras}.
\par
Finally, we investigated the general properties of timelike as well as null geodesics for the wormhole solutions we found. In the case of null geodesics, gravitational lensing effects on the wormhole's surrounding environment were studied and it was found that the effective potential of photons traveling in the wormhole spacetime admits a maximum value at the throat. This result was obtained for a subclass of zero and nonzero redshift solutions and one may be temped to find more complicated behavior for the effective potential, at or outside of the wormhole throat, see e.g.,~\cite{nisha}. This depends on whether the proper radial distance, specifically for non-constant redshift solutions Eq.~(\ref{bsolfinal}), can be exactly obtained in terms of standard mathematical functions. However, the resultant integral is too complicated to be solved analytically and more advanced mathematical techniques are needed in order to overcome the problem. Consequently, for this subclass of solutions we obtained deflection angle of the incoming beam of light admits positive values and diverges in the limit of approach to the throat. Therefore the wormhole configuration can act as a converging lens with photon sphere located at the throat. The effective potential for timelike geodesics was calculated for both zero and nonzero tidal force solutions. In the former we observed that the effective potential admits a global maximum at the throat which can discriminate between the particles passing through the throat from upper to lower Universes and those that reflected back by the potential barrier. In the latter case, we found that the effective potential can admit a maximum at the throat and a local minimum beyond it. Hence, depending on total energy of the particle we may have, $i)$ trajectories in which the particle can move through the throat of wormhole from the upper Universe to the lower one, $ii)$ trajectories in which the particle is reflected back by the potential barrier and remains in the very Universe and, $iii)$ trajectories in which the particle can move on oscillatory paths subject to bound orbits in one of the upper/lower Universes. Further discussions on particle motion in wormhole spacetime can be found in~\cite{parttrwosp}.


\begin{thebibliography}{99}
\bibitem{Flamm} L. Flamm, Phys. Z. {\bf 17} 448 (1916); G. W. Gibbons, Gen. Relativ. Gravit. {\bf47} 72 (2015).
\bibitem{Weyl1920} H. Weyl, \lq{}\lq{}{\it Philosophy of Mathematics and Natural Science,}\rq{}\rq{} Princeton University, Princeton, NJ, (1949).
\bibitem{ERBR} A. Einstein and N. Rosen, Phys. Rev. {\bf 48}, 73 (1935);\\
D. R. Brill and R. W. Lindquist, Phys. Rev. {\bf 131}, 471 (1963).
\bibitem{Wheel1972} J. A. Wheeler, Phys. Rev. {\bf 97}, 511 (1955).
\bibitem{MisWheel} C. W. Misner and J. A. Wheeler, Ann. Phys. {\bf 2}, 525 (1957); \\C. W. Misner, Phys. Rev. {\bf 118}, 1110 (1960).
\bibitem{Wheelerworm} J. A. Wheeler, Ann. Phys. {\bf 2}, 604 (1957);\\ J. A. Wheeler, \lq{}\lq{}{\it Geometrodynamics,}\rq{}\rq{} Academic, New York, (1962).
\bibitem{FulWheel} R. W. Fuller and J. A. Wheeler, Phys. Rev. {\bf 128}, 919 (1962).
\bibitem{mt} M. S. Morris and K. S. Thorne, Am. J. Phys. {\bf 56}, 395
(1988).
\bibitem{mt1} M. S. Morris, K. S. Thorne and U. Yurtsever, Phys. Rev. Lett. {\bf 61}, 1446 (1988).
\bibitem{khu} M. Visser, \lq{}\lq{}{\it Lorentzian Wormholes: From Einstein to Hawking,}\rq{}\rq{} AIP, Woodbury, USA, (1995);\\D. Hochberg and M. Visser, Phys. Rev. D {\bf 56}, 4745 (1997).
\bibitem{ghostsfph} C. Armendariz-Picon, Phys. Rev. D {\bf 65}, 104010 (2002);\\S. V. Sushkov, Phys. Rev. D {\bf 71}, 043520 (2005);\\F. S. N. Lobo, Phys. Rev. D {\bf 71}, 084011 (2005).
\bibitem{timemach} O. Bertolami and R. Z. Ferreira, Phys. Rev. D {\bf 85}, 104050 (2012);\\V. P. Frolov and I. D. Novikov, Phys. Rev. D {\bf 42}, 1057 (1990);\\M. Visser, Phys. Rev. D {\bf 47}, 554 (1993).
\bibitem{HawCMF} S. W. Hawking, G. F. R. Ellis, \lq{}\lq{}{\it The large scale structure of spacetime,}\rq{}\rq{} Cambridge University Press, Cambridge, (1973).
\bibitem{Caseffect} H. Epstein, V. Glaser and A. Jaffe, IL Nuovo Cimento, {\bf 36}, 1016 (1965).
\bibitem{negendensqueez} D. Hochberg, T. W. Kephart, Phys. Lett. B {\bf 268}, 377 (1991).
\bibitem{Klinkhammer1991} G. Klinkhammer, Phys. Rev. D {\bf 43}, 2542 (1991).
\bibitem{phantworm} F. S. N. Lobo, Phys. Rev. D {\bf 71}, 124022 (2005); \\P. K. F. Kuhfittig, Class. Quant. Grav. {\bf 23}, 5853 (2006); \\F. S. N. Lobo, F. Parsaei and N. Riazi, Phys. Rev. D {\bf 87}, 084030 (2013);\\
Y. Heydarzade, N. Riazi and H. Moradpour, Can. J. Phys. {\bf 93}, 1523 (2015).
\bibitem{csworm} R. Garattini, Eur. Phys. J. C {\bf 79}, 951 (2019);\\Z. Hassan, S. Ghosh, P. K. Sahoo, V. S. H. Rao, Gen. Relativ. Grav. {\bf 55}, 90 (2023);\\O. Sokoliuk, A. Baransky, P. K. Sahoo, Nuc. Phys. B, {\bf 930} 115845 (2022);\\S. K. Tripathy, Phys. Dark Univ., {\bf 31}, (2021) 100757;\\R. Garattini, Eur. Phys. J. C {\bf 81}, 824 (2021);\\
P. H. F. Oliveira, G. Alencar, I. C. Jardim, R. R. Landim, Symmetry {\bf 15}, 383 (2023); R. Avalos, E. Fuenmayor, E. Contreras, Eur. Phys. J. C, {\bf 82}, 420 (2022);\\A. H. Ziaie, M. R. Mehdizadeh, Class. Quantum Grav. {\bf 41}, 145001 (2024).
\bibitem{intdarksec} V. Folomeev and V. Dzhunushaliev, Phys. Rev. D {\bf 89}, 064002 (2014).
\bibitem{lobocgqreview} F. S. N. Lobo, Classical and Quantum Gravity Research, 1-78, (2008), Nova Sci. Pub. ISBN 978-1-60456-366-5, arXiv:0710.4474 [gr-qc];\\ F. S. N. Lobo, Int. J. Mod. Phys. D {\bf 25}, 1630017 (2016).
\bibitem{bd} A. G. Agnese and M. La Camera, Phys. Rev. D {\bf 51}, 2011 (1995);\\
K. K. Nandi, A. Islam, and J. Evans, Phys. Rev. D {\bf 55}, 2497 (1997);\\  L. A. Anchordoqui, S. P. Bergliaffa, and D. F. Torres, Phys. Rev. D {\bf 55}, 5226 (1997);\\R. Shaikh and S. Kar, Phys. Rev. D {\bf 94}, 024011 (2016).
\bibitem{Garcia-Lobo} N. M. Garcia and F. S. N. Lobo, Phys. Rev. D {\bf 82}, 104018 (2010);\\ M. Zubair, S. Waheed and Y. Ahmad, Eur. Phys. J. C {\bf 76}, 444 (2016);\\
S. K. Tripathy, Phys. Dark Univ., {\bf 31}, 100757 (2021);\\A. Dixit, C. Chawla, A. Pradhan, Int. J. Geom. Method Mod. Phys. {\bf 18}, 2150064 (2021);\\ R. Solanki, Z. Hassan, P. K. Sahoo, Chinese J. Phys., {\bf 85} 74 (2023);\\ P. H. R. S. Moraes, A. S. Agrawal, B. Mishra, Phys. Lett. B, {\bf 855}, 138818 (2024);\\ L. V. Jaybhaye, M. Tayde and P. K. Sahoo, Commun. Theor. Phys. {\bf 76}, 055402 (2024).
\bibitem{LOVEWORM} G. Dotti, J. Oliva, R. Troncoso, Phys. Rev. D {\bf 75}, 024002 (2007);\\
H. Maeda, M. Nozawa, Phys. Rev. D {\bf 78}, 024005 (2008);\\M. H. Dehghani and Z. Dayyani, Phys. Rev. D {\bf 79}, 064010 (2009);\\M. R. Mehdizadeh and F. S. N. Lobo,  Phys. Rev. D {\bf 93}, 124014 (2016).
\bibitem{fr} N. Furey and A. DeBenedictis, Class. Quantum Grav. {\bf 22}, 313 (2005);\\ F. S. N. Lobo and M. A. Oliveira, Phys. Rev. D {\bf 80}, 104012 (2009); \\A. De Benedictis, D. Horvat, Gen. Relat. Gravit. {\bf 44}, 2711 (2012);\\
M. Sharif and I. Nawazish, Annals of Physics, {\bf 389}, 283 (2018);\\ R. Radhakrishnan, P. Brown, J. Matulevich, E. Davis, D. Mirfendereski, G. Cleaver, Symmetry, {\bf 16}, 1007 (2024).
\bibitem{ectwormhole1} K. A. Bronnikov and A. M. Galiakhmetov, Grav. Cosmol. {\bf 21}, 283 (2015).
\bibitem{ectwormhole} M. R. Mehdizadeh and A. H. Ziaie, Phys. Rev. D {\bf 95}, 064049 (2017);\\ Phys. Rev. D {\bf 99}, 064033 (2019).
\bibitem{gmfl} S. H. Mazharimousavi, M. Halilsoy, and Z. Amirabi, Phys.
Rev. D {\bf 81}, 104002 (2010);\\ P. Kanti, B. Kleihaus and J. Kunz, Phys. Rev. D {\bf 85}, 044007 (2012);\\
G. Antoniou, A. Bakopoulos, P. Kanti, B. Kleihaus and Jutta Kunz, arXiv:1904.13091 [hep-th].
\bibitem{branww} F. Parsaei and N. Riazi, Phys. Rev. D {\bf 102}, 044003 (2020);\\Y. Akai and K.-ichi Nakao, Phys. Rev. D {\bf 96}, 024033 (2017);\\Y. Tomikawa, T. Shiromizu, K. Izumi, Phys. Rev. D {\bf 90}, 126001 (2014).
\bibitem{otherworms} R. Shaikh, Phys. Rev. D {\bf 92}, 024015 (2015);\\ F. Rahaman, N. Paul, A. Banerjee, S. S. De, S. Ray and A. A. Usmani, Eur. Phys. J. C {\bf 76}, 246 (2016);\\ P. H. R. S. Moraes, P. K. Sahoo, Phys. Rev. D {\bf 96}, 044038 (2017);\\ M. G. Richarte, I. G. Salako, J. P. Morais Graca, H. Moradpour, and A. ovgun, Phys. Rev. D {\bf 96}, 084022 (2017);\\K. Jusufi, N. Sarkar, F. Rahaman, A. Banerjee and S. Hansraj, Eur. Phys. J. C {\bf 78} 349 (2018);\\ V. De Falco, E. Battista, S. Capozziello, M. De Laurentis, Eur. Phys. J. C {\bf 81}, 157 (2021);\\ V. De Falco, E. Battista, S. Capozziello, M. De Laurentis, Phys. Rev. D {\bf 103}, 044007 (2021);\\ J. L. Rosa, J. P. S. Lemos, F. S. N. Lobo, Phys. Rev. D {\bf 98}, 064054 (2018).
\bibitem{wormrastrace} G. Mustafa, M. R. Shahzad, G. Abbas, and T. Xia, Mod. Phys. Lett. A {\bf 33} 2050035 (2020).
\bibitem{morad1} H. Moradpour, N. Sadeghnezhad, S. H. Hendi, Can. J. Phys, {\bf 95} 1257 ()2017).
\bibitem{morad2} S. Halder, S. Bhattacharya, S. Chakraborty, Mod. Phys. Lett. A {\bf 34}, 1950095 (2019).
\bibitem{ras} P. Rastall, Phys. Rev. D {\bf 6}, 3357 (1972);\\
P. Rastall, Can. J. Phys. {\bf 54}, 66 (1976).
\bibitem{morad3} H. Moradpour, Y. Heydarzade, F. Darabi and Ines G. Salako, Eur. Phys. J. C {\bf 77}, 259 (2017).
\bibitem{ppp} G. W. Gibbons and S. W. Hawking, Phys. Rev. D {\bf 15}, 2738 (1977);\\
N. D. Birrell and P. C. W. Davies, \lq{}\lq{}{\it Quantum Fields in Curved Space,}\rq{}\rq{} Cambridge University Press, Cambridge, (1982).
\bibitem{ppp1} L. Parker, Phys. Rev. D {\bf 3}, 346 (1971), [Erratum: Phys. Rev.D {\bf 3}, 2546 (1971)].
\bibitem{rascosastro} H. Moradpour, A. Bonilla, E. M. C. Abreu, and J. A. Neto, Phys. Rev. D {\bf 96}, 123504 (2017);\\H. Moradpour, Y. Heydarzade, C. Corda, A. H. Ziaie,
S. Ghaffari, Mod. Phys. Lett. A, {\bf 33}, 1950304 (2019);\\ F.-F. Yuan and P. Huang, Class. Quant. Grav. {\bf 34}, 077001 (2017);\\I. P. Lobo, H. Moradpour, J. P. Morais Graca, and I. G. Salako, Int. J. Mod. Phys. D {\bf 27}, 1850069 (2018);\\R. Kumar, S. G. Ghosh, Eur. Phys. J. C {\bf 78}, 750 (2018);\\K. Bamba, A. Jawad, S. Rafique, et al., Eur. Phys. J. C {\bf 78}, 986 (2018);\\H. Moradpour and M. Valipour, Can. J. Phys. {\bf 98}, 853 (2020);\\ S. Halder, S. Bhattacharya, S. Chakraborty, Mod. Phys. Lett. A {\bf 34}, 1950095 (2019);\\R. Li, J. Wang, Z. Xu and X. Guo, MNRAS, {\bf 486}, 2407 (2019);\\A. M. Oliveira, H. E. S. Velten, J. C. Fabris, L. Casarini, Phys. Rev. D {\bf 92}, 044020 (2015);\\ S. K. Maurya, F. Tello-Ortiz, Phys. Dark Univ. {\bf 29}, 100577 (2020);\\X.-C. Cai, Y.-G. Miao, Phys. Rev. D {\bf 101}, 104023 (2020).
\bibitem{morad4} F. Darabi, H. Moradpour, I. Licata, et al., Eur. Phys. J. C {\bf 78}, 25 (2018).
\bibitem{GRG22} D. Das, S. Dutta and S. Chakraborty, Eur. Phys. J. C {\bf 78}, 810 (2018).
\bibitem{ZiaRas} A. H. Ziaie, H. Moradpour, M. Mohammadi Sabet, Eur. Phys. J. Plus {\bf 136}, 1085 (2021).
\bibitem{rascosmiceras1} C. E. M. Batista, M. H. Daouda, J. C. Fabris, O. F. Piattella and D. C. Rodrigues, Phys. Rev. D {\bf 85}, 084008 (2012).
\bibitem{Lemos2003} J. P. S. Lemos, F. S. N. Lobo, S. Q. de Oliveira, Phys. Rev. D {\bf 68}, 064004 (2003).
\bibitem{FLoboBook} F. S. N. Lobo (Editor), \lq{}\lq{}{\it Wormholes, Warp Drives and Energy Conditions,}\rq{}\rq{} Springer (2017).
\bibitem{Hobs} M. P. Hobson, G. P. Efstathiou, A. N. Lasenby, \lq{}\lq{}{\it General Relativity: An Introduction for Physicists,}\rq{}\rq{} United Kingdom, Cambridge University Press, (2006).
\bibitem{wormbhtsu} N. Tsukamoto, T. Harada, K. Yajima, Phys Rev. D {\bf 86}, 104062 (2012).
\bibitem{SWeinbergbook} S. Weinberg, \lq{}\lq{}{\it Gravitation and cosmology: principles and applications of the general theory of relativity,}\rq{}\rq{} Wiley (1972);\\V. Bozza, Gen. Relativ. Gravit. {\bf 42}, 2269 (2010);\\R. Shaikh, P. Banerjee, S. Paul, T. Sarkar, Phys. Rev. D {\bf 99}, 104040 (2019).
\bibitem{Bozza} V. Bozza, Phys. Rev. D {\bf 66}, 103001 (2002).
\bibitem{Hasse-Perlick} W. Hasse and V. Perlick, Gen. Relativ. Gravit. {\bf 34}, 415 (2002).
\bibitem{Perlicklvr} V. Perlick, Living Rev. Relativity, {\bf 7}, 9 (2004);\\J. H. Simonetti, M. J. Kavic, D. Minic, D. Stojkovic, D.-Chang Dai, Phys. Rev. D {\bf 104}, 081502 (2021);\\ C. Bambi, D. Stojkovic, Universe, 7(5), 136 (2021).
\bibitem{Shaikh-novel} R. Shaikh, P. Banerjee, S. Paul and T. Sarkar, Phys. Lett. B {\bf 789}, 270 (2019).
\bibitem{highsensinstru} K. S. Virbhadra, G. F. R. Ellis, Phys. Rev. D {\bf 62}, 084003 (2000);\\ M. Silvia and R. Esteban, \lq{}\lq{}{\it Gravitational Lensing And Microlensing,}\rq{}\rq{} World Scientific (2002);\\ F. Courbin and D. Minniti, (Eds.), \lq{}\lq{}{\it Gravitational Lensing: An Astrophysical Tool,}\rq{}\rq{} Springer (2008);\\ M. Kilbinger, Rep. Prog. Phys. {\bf 78}, 086901 (2015);\\ S. Dodelson, \lq{}\lq{}{\it Gravitational Lensing,}\rq{}\rq{} Cambridge University Press (2017).
\bibitem{wormmodif} F. S. N. Lobo, M. A. Oliveira, Phys. Rev. D {\bf 80}, 104012 (2009);\\A. Chanda, S. Dey, B. C. Paul, Gen. Relativ. Gravit. {\bf 53}, 78 (2021);\\T. Harko, F. S. N. Lobo, M. K. Mak, S. V. Sushkov, Phys. Rev. D {\bf 87}, 067504 (2013);\\ S. Bhattacharya, S. Halder, S. Chakraborty, Mod. Phys. Lett. A {\bf 34}, 1950200 (2019).
\bibitem{moraddyn} H. Shabani, H. Moradpour, A. H. Ziaie, Phys. Dark Univ., {\bf 36}, 101047 (2022).
\bibitem{Lin2020} K. Lin, WL. Qian, Eur. Phys. J. C {\bf 80}, 561 (2020). 
\bibitem{struformgrg} A. H. Ziaie, H. Moradpour, H. Shabani, Eur. Phys. J. Plus {\bf 135}, 916 (2020).
\bibitem{shaban} H. Shabani, A. H. Ziaie, H. Moradpour, Ann. Phys., {\bf 444}, 169058 (2022).
\bibitem{regbhras} K. Lin and W.-L. Qian, Chinese Phys. C {\bf 43}, 083106 (2019).
\bibitem{nisha} R. Shaikh, P. Banerjee, S. Paul, T. Sarkar, JCAP 07 (2019) 028;\\ N. Godani, Int. J. Geom. Meth. Mod. Phys. {\bf 20}, 2350005 (2023);\\N. Godani, G. C. Samanta, Ann. Phys., {\bf 429}, 168460 (2021);\\ N. Tsukamoto and T. Harada, Phys. Rev. D {\bf 95}, 024030 (2017).
\bibitem{parttrwosp} M. Cataldo, L. Liempi, P. Rodriguez, Eur. Phys. J. C, {\bf 77}, 748 (2017);\\TY. Zhou, and Y. Xie, Eur. Phys. J. C {\bf 80}, 1070 (2020);\\F. Abdulxamidov, C. A. B.-Gallego, W.-B. Han, J. Rayimbaev, A. Abdujabbarov, Phys. Rev. D {\bf 106} 024012 (2022);\\ C. A. B.-Gallego, W.-B. Han, D. Malafarina, B. Ahmedov, A. Abdujabbarov, 	Phys. Rev. D {\bf 104}, 084024 (2021);\\A. Dutta, D. Roy, S. Chakraborty, New Astron. {\bf 111}, 102236 (2024).
\end{thebibliography}
\end{document}